\documentclass[doublecol]{epl2} 

\usepackage{graphicx}
\newcommand{\mm}{\mathrm}

\title{Determination of the universality class of crystal plasticity}
\date{\today}

\author{G. Tsekenis\inst{1} \and J.T. Uhl \and N.Goldenfeld\inst{1} \and K. A. Dahmen\inst{1} }
\shortauthor{G. Tsekenis \etal}

\institute{                    
  \inst{1} Department of Physics, University of Illinois at Urbana-Champaign, Loomis Laboratory of Physics - 1110 West Green
Street, Urbana, Illinois, 61801-3080.
}

\abstract{
Although scaling phenomena have long been documented in crystalline
plasticity, the universality class has been difficult to identify due
to the rarity of avalanche events, which require large system sizes and
long times in order to accurately measure scaling exponents and functions.
Here we present comprehensive simulations of two-dimensional
dislocation dynamics under shear, using finite-size scaling to extract
scaling exponents and the avalanche profile scaling function from
time-resolved measurements of slip-avalanches. Our results provide
compelling evidence that both the static and dynamic universality
classes are consistent with the mean-field interface depinning model.}

\begin{document}

\maketitle

\section{Introduction}
Crystalline materials deform in a plastic, irreversible manner at sufficiently high stresses. Bulk continuum theories successfully
reproduce several macroscopic features of plastic flow such as the stress-strain curve and work-hardening~\cite{hosford2005mbm}. This success is mainly due to the fluctuations averaging out at macroscopic scales and therefore deformation appears to be smooth in time and homogeneous in space.

At microscopic scales crystal deformation is both spatially
inhomogeneous and intermittent in time. Topological defects such as
dislocations move intermittently, causing the material to slip in
discrete steps. Those defects interact with each other via long-range
elastic interactions, mediated through the material and respond
collectively to external stresses, giving slip avalanches. These slip
avalanches are characterized by long-range correlations in space and
time giving avalanche sizes distributed according to power-laws for
several orders of magnitude~\cite{MiguelNat01, WeissMSEA01,
RichetonNatMat05, Richeton06, Richeton05, WeissJGR00, Weiss97,
weiss2003sfa, DimidukSci06, ZaiserAdvPhys06}.

Despite intense computational efforts to predict a complete set of
universal (i.e., detail-independent) power-law exponents, there is an
ongoing debate about their values and the corresponding universality
class of systems that share the same exponents. Several previous
discrete dislocation dynamics simulations have reported contradictory
results for static and dynamic power law exponents
~\cite{ZaiserAdvPhys06,Laurson06,MiguelPRL02}.  (We call properties
\lq\lq dynamic"  if they resolve the dynamics {\it during} the propagation of
an individual avalanche, and \lq\lq static" if they do not.) Here we present
a consistent picture  that strongly supports the claim that   both the
statics and the dynamics of crystal plasticity simulations agree  with
mean-field theory predictions~\cite{ZaiserAdvPhys06,UhlPRL09}, and
therefore they both belong to the mean-field interface depinning
universality class of all systems that share the same exponents.
Knowing the values of these exponents is important for applications.
For example the dynamical depinning exponent $\beta$
~\cite{MiguelPRL02} describes how quickly a crystal deforms as a
function of stress, and the power spectra~\cite{Laurson06} can be used
to obtain information about the deformation mechanism and material
failure from non-destructive acoustic emission experiments. Also, the
power law exponents do not depend on material details, so they are an
ideal quantity for testing the predictions of the simple coarse grained
models against experiments.

%\begingroup
\begin{table*}[!htbp]
\begin{center}
\caption{Table of exponents. Our results from 2D DDD are in the second column.
The exponents we extracted directly from our simulations are under \lq \lq extracted"
while the exponents we derived through exponent relations (indirectly) are under \lq \lq derived".
%The error bars on exponents that were extracted from a scaling collapse indicate
%a range of values that gave a similar good quality collapse.
Results from literature from full 3D DDD are indicated with * and
from 2D DDD with creation and annihilation in the steady state with $^{+}$.
In the numerical work of Refs.~\cite{ZaiserYielding,ZaiserAdvPhys06}, the total slip of the dislocation system
$L_{\mm{aval}}=\int_{T}dt\sum^{N}_{i=1}b_{i}v_{i}(t)=\sum^{N}_{i=1}b_{i}\Delta
x_{i,T}$ was used to measure the size of the avalanche. Our simulations
calculate the collective slip produced by the dislocation system
$S=\int_{T}dt\sum^{N}_{i=1}|v_{i}(t)|=\sum^{N}_{i=1}|\Delta
x_{i,T}|$ during an avalanche. For large avalanches, the total slip and
collective slip have the same scaling behavior.
%or the maximum peak activity squared, $V_{\mm{max}}^{2}$~\cite{Richeton05,MiguelNat01} ($E$ was used in~\cite{Richeton06}).
%(It is  common in acoustic emission experiments to consider the distribution of the maximum peak activity, $V_{\mm{max}}$, during an avalanche, e.g.~\cite{WeissMSEA01,Richeton05,WeissJGR00,Weiss97}. However extremal (a.k.a. order) statistics for non-independent and non-identically distributed variables (non-i.i.d.) are complicated to calculate analytically. The values $V(t_{i})$ in an avalanche are correlated because the power spectrum is not white noise. Therefore the derivation of the scaling of the statistics of $V_{\mm{max}}$
%and $V_{\mm{max}}^2$
%is not straightforward. On the other hand the distributions of the avalanche measures $S$, $E$ and $L_{\mm{aval}}$ are easy to derive and compare to MFT. In any case the power law distribution of $V_{\mm{max}}$ is useful knowledge although it comes with large errorbars.)
}

\begin{tabular}{c|c|c|c|c}
\multicolumn{1}{c}{exponent}&\multicolumn{1}{c}{extracted}&\multicolumn{1}{c}{MFT}& \multicolumn{1}{c}{simulations}&\multicolumn{1}{c}{experiments}\\
\hline \hline
$\kappa$      & $1.5\pm0.1$ & $\frac{3}{2}$ & 1.4~\cite{ZaiserYielding},1.6~\cite{MiguelMSEA01},1.5~\cite{CsikorSci07}* & 1.5-1.6~\cite{DimidukSci06},1.5~\cite{GreerPRL08}\\
%"      &  &  & 1.5~\cite{CsikorSci07}* & \\
$\frac{1}{\sigma}$  & $2\pm0.2$ & 2 & 2~\cite{ZaiserYielding},2~\cite{CsikorSci07}* & 2~\cite{ZaiserYielding} \\
\hline
$1+\frac{\kappa-1}{2-\sigma \nu z}$  & $1.3\pm0.1$ & $\frac{4}{3}$ & $1.8 \pm0.2$~\cite{MiguelNat01}$^{+}$ & $1.5\pm0.1$~\cite{Richeton06},$1.6\pm0.05$~\cite{MiguelNat01}\\
% " & & & & $1.6\pm0.05$~\cite{MiguelNat01}\\
$\frac{2-\sigma \nu z}{\sigma}$   & $3\pm0.3$ & 3 &  & \\
\hline
$\langle S \rangle \sim T^{\frac{1}{\sigma \nu z}}$   & $2\pm0.2$ & 2 & 1.5~\cite{Laurson06}$^{+}$ & \\
$\langle T \rangle \sim S^{\sigma \nu z}$   & $0.5\pm0.1$ & $\frac{1}{2}$ & & \\
\hline
$\frac{1}{\sigma \nu z}$ (fig.~\ref{PS}) & $2\pm0.1$ &  2 & 1.5~\cite{Laurson06}$^{+}$ & \\
$\frac{1}{\sigma \nu z}$ (fig.~\ref{AvalShapes})  & $1.9\pm0.1$ & 2 & 1.5~\cite{Laurson06}$^{+}$ &\\
\hline 
$\nu$  (fig.~\ref{vtau}) & $1\pm0.2$ & 1 &  & \\
\hline
$\beta$ (fig.~\ref{vtau})  & $1.17\pm0.02$ & 1 & $1.8\pm0.1$~\cite{MiguelPRL02}$^{+}$ & \\
\hline \hline
\multicolumn{1}{c}{}&\multicolumn{1}{c}{derived}\\
\hline
$1+\frac{\kappa-1}{\sigma \nu z}$     & $2\pm0.2$  & 2 & & \\
$\nu z$                                                    & $1\pm0.1$    & 1 & & \\
\hline
$z$                                & $1\pm0.2$ & 1 &  & \\
\hline
$\beta$                         & $1\pm0.25$ & 1 & $1.8\pm0.1$~\cite{MiguelPRL02}$^{+}$ & \\
\hline \hline
\end{tabular}
\label{table}
\end{center}
\end{table*}
%\endgroup

%\clearpage

A simple analytic mean-field theory (MFT)~\cite{ZaiserAdvPhys06,UhlPRL09} for plasticity
suggests that the observed power law scaling of the slip-avalanche size
distributions is the reflection of an underlying non-equilibrium
critical point~\cite{Sethna01}, which is located at the critical flow
stress. The critical flow stress $\tau_c$ separates a low stress phase
where the material can sustain loads on the time scales of typical
experiments, from a high stress phase where the material \lq\lq flows",
or \lq\lq gives way" by deforming continually under loads that are
higher than the critical flow stress. Below the critical flow stress
$\tau_c$, at fixed stress $\tau<\tau_c$, the average strain rate is
zero and dislocations are stuck on average, while above the critical
flow stress they move continually, and the average strain rate is
nonzero.  For stresses $\tau>\tau_c$ the strain rate $d\gamma/dt$
scales as $d\gamma/dt \sim (\tau-\tau_c)^\beta$ where $\beta$ is the
depinning exponent~\cite{MiguelPRL02}. In mean-field theory $\beta=1$.
Below the critical flow stress, when the stress is increased by a small
step, the system responds with a dislocation slip avalanche, at the end
of which all dislocations are re-pinned again and remain stuck until
the stress is increased again. As the stress slowly approaches the
critical flow stress from below, the average slip avalanche size
$\langle s \rangle$ grows bigger and it diverges at the critical point
as $\langle s \rangle \sim (\tau_c-\tau)^{(\kappa-2)/\sigma}$, where
$\kappa =1.5$, and $\sigma =1/2$ in mean-field theory.  The average
avalanche size at a fixed stress can thus be used as a measure of the
proximity to the critical flow stress.

The purpose of this Letter is to provide the first comprehensive
calculation of the time-resolved behavior of slip avalanches.  We use
finite-size scaling to compute accurately a full suite of critical
exponents and the associated scaling function, in order to determine
the static and dynamic universality class. There are many additional
critical exponents that are predicted by MFT~\cite{UhlPRL09} (see Table
~\ref{table}) and they all belong to the mean-field  interface depinning
universality class~\cite{ZaiserAdvPhys06,UhlPRL09}. Analyses using
renormalization group techniques suggest that the interaction range of
dislocations is sufficiently long range so that mean-field theory,
which uses infinite range interactions, predicts the correct exponents
for 2- and 3-dimensional crystals~\cite{ZaiserAdvPhys06,UhlPRL09}.  The
theoretical expectations have in the past not been confirmed by
simulations, and it is this inconsistency that we address here.

Discrete dislocation dynamics models~\cite{MiguelNat01, ZaiserYielding,
MiguelMSEA01, Laurson06, CsikorSci07, Weygand2010, TsekenisJam11},
continuum models~\cite{ZaiserYielding}, phase field models
~\cite{Koslowski} and phase field crystal models~\cite{chanPRL10}
indicate a nonequilibrium critical point, but no consensus has been
reached on its universality class. Zaiser~\cite{ZaiserYielding}
achieved a scaling collapse of static properties, such as the simulated
slip-avalanche size distribution at different external stresses below
the critical flow stress, with critical exponents that are consistent
with mean-field theory. Other discrete dislocation dynamics simulations
obtained dynamic quantities that did not agree with mean-field theory:
Laurson et al~\cite{Laurson06} reported that the power spectra of the
slip-velocity time series above the flow stress are characterized by a
critical exponent that differs from the mean-field theory predictions.
Miguel et al~\cite{MiguelPRL02} found an independent depinning exponent
$\beta$ that also differs from mean-field theory predictions.

Part of the difficulty in resolving the differences between the static
and the dynamic results is that the long range interactions lead to
unusually prominent finite-size effects that can skew numerical scaling
results. To circumvent these finite-size effects we perform a finite
size scaling analysis of the avalanche statistics obtained from our
discrete dislocation dynamics simulations. We calculate both universal
scaling exponents and universal scaling functions associated with the
\textit{temporal} profiles of the slip speed during avalanches. We find
that the power spectra of the slip speed time series (below and above
the critical flow stress) exhibit power-law behavior and that the
avalanche shapes collapse with matching exponents. More importantly, we
find that both the collapse function and a comprehensive set of 13
exponents obtained from our simulations (Table~\ref{table}) for both
static and dynamic properties, are in excellent agreement with the
simple model~\cite{UhlPRL09} in the mean-field interface depinning
universality class, including the finite-size scaling exponent $\nu$
and the depinning exponent $\beta$. Our work thus demonstrates that
even though there is no apparent quenched disorder in these systems,
the time-resolved and finite-size scaling properties of the dislocation
system behave according to the mean-field interface depinning model
which does have quenched disorder.

Our work makes quantitative predictions for the scaling behavior of
dislocation systems at sufficiently large length- and long time-scales where
the microscopic details should not be important. Therefore 
our scaling results are relevant to the deformation of micro-
~\cite{DimidukSci06, MiguelNat01} and nano-pillars~\cite{GreerPRL08,
GreerJenningsPRL10} alike, for pillars that are large enough to display
collective dislocation dynamics. Recent experimental studies on
nanopillars confirm these predictions~\cite{NirnanoPRL12}.

\section{Discrete dislocation dynamics model} In order to study the
avalanches of plasticity we employ discrete dislocation dynamics (DDD)
simulations in two dimensions (2D). The details of our model can be
found in~\cite{TsekenisJam11}. They are similar to other 2D DDD models
in the literature~\cite{MiguelNat01, ZaiserYielding, Laurson06,
Weygand2010}. In brief, in a square box of side $L$, we place $N$
straight edge dislocations parallel to the $z$-axis. The dislocations
are allowed to move continuously along the $x$-axis, the shear
direction, while their $y$ position is fixed. Each dislocation is
assigned a Burgers vector $\vec{b_{i}}=\pm \hat{x}$ such that
$\sum^{N}_{i=1}{b_{i}}=0$. Every pair at a distance $\vec{r}=(x,y)$
interacts via the interaction stress $\tau_{\mm{int}}$,
\begin{eqnarray}
\tau_{\mm{int}}(\vec{r})=\frac{b\mu}{2\pi(1-\nu)} \frac{x(x^2-y^2)}{(x^2+y^2)^2}
\label{tauint}
\end{eqnarray}
where $\mu$ is the shear modulus and $\nu$ is the Poisson ratio of the
host medium. Each dislocation moves in response to $\tau_{\mm{int}}$,
and the external shear stress $\tau_{\mm{ext}}(\equiv \tau)$. Their stick-slip motion can be
described by overdamped equations of motion:
\begin{eqnarray}
\eta \frac{dx_{i}}{dt} = b_{i} \left( \sum^{N}_{j \neq i} \tau_{\mm{int}}(\vec{r}_{j}-\vec{r}_{i})+\tau_{\mm{ext}} \right)
\end{eqnarray}
for $ i,j=1,\ldots,N$ where $x_{i}$ is the $x$ coordinate of the $i$th
dislocation at point $\vec{r}_{i}$ with Burgers vector $b_i$, $t$ is
time and $\eta$ is the effective viscosity in the host medium
~\cite{MiguelNat01,Laurson06,ZaiserYielding}.
($\sum^{N}_{j \neq i} \tau_{\mm{int}}(\vec{r}_{j}-\vec{r}_{i})$ is a dynamically changing
inhomogeneous stress field which pins the dislocations for $\tau<\tau_c$.)
In our computer
simulations we set the temperature to zero, the distance scale to $b=1$
and the time scale to $t_{0}=\eta /(\mu/(2\pi(1-\nu))=1$. We impose
periodic boundary conditions in both $x$ and $y$ directions and use the
Lekner summation method~\cite{Lekner} of image cells to treat the
long-range character of the dislocation interaction.
The choice of the boundary conditions does not affect
the scaling behavior on long length scales~\cite{Sethna01, NirnanoPRL12, GreerPRL08}.
Neither does creation and annihilation affect the power law exponents
and the scaling functions presented here~\cite{ZaiserAdvPhys06, UhlPRL09}.

We solve the equations of motion with the adaptive-step fifth-order Runge-Kutta
method~\cite{NumRecC}. We keep the dislocation number constant, since
we do not want to consider dislocation creation or annihilation. We
define the dislocation collective speed (also called activity) as,
\begin{eqnarray}
V(t)=\sum^{N}_{i=1}|v_{i}(t)|
\end{eqnarray}
where $v_{i}=dx_{i}/dt$. The acoustic emission signal is proportional
to $V(t)$. Another choice is $V'(t)=\sum^{N}_{i=1}b_{i}v_{i}(t)$, which
is proportional to the strain rate~\cite{ZaiserAdvPhys06}. The
avalanches produced from either of these two measures converge to the
same scaling behavior for large avalanches.

\section{Below the critical flow stress} We start by  randomly seeding the $N$
dislocations in the simulation box and letting the system relax to the
nearest (metastable) equilibrium state at zero external stress. The
dislocation activity approaches zero as the system approaches the
nearest local energy minimum. A simple eigenmode analysis shows that
the time needed for the system to reach zero activity diverges. When
the dislocation activity has fallen below a threshold the system is
sufficiently close to the energy minimum. We increase the external
stress adiabatically (or quasi-statically) slowly whenever and for as
long as the system's activity is below the specified threshold,
$V(t)<V_{\mm{th}}$. Eventually the increased external stress pushes the
system's activity above the threshold (this is the starting time of an
avalanche $t_{\mm{start}}$). During the time that $V(t)>V_{\mm{th}}$
the system produces an avalanche and we keep the external stress
constant until the avalanche has completed and the activity falls below
threshold (this is the ending time of an avalanche $t_{\mm{end}}$); the
avalanches do not overlap in time.

For relatively low values of the external stress the system responds
with small avalanches. As the stress $\tau$ approaches the critical flow stress
$\tau_c$, it responds with larger and larger avalanches until at
$\tau_c$ it finally flows steadily with an infinite avalanche. When the
applied stress exceeds the critical value, i.e., $\tau > \tau_c$, we
observe the dislocations moving continually, exiting from one side of
the simulation cell and reemerging at the other due to the periodic
boundary conditions, without ever getting stuck again. This is the
point when the sample flows in a deformation experiment. In summary,
for $\tau < \tau_c$ the system is pinned. For $\tau > \tau_c$ the
system is flowing.

\begin{figure}[h]
\centering
    \includegraphics[width=0.49\textwidth]{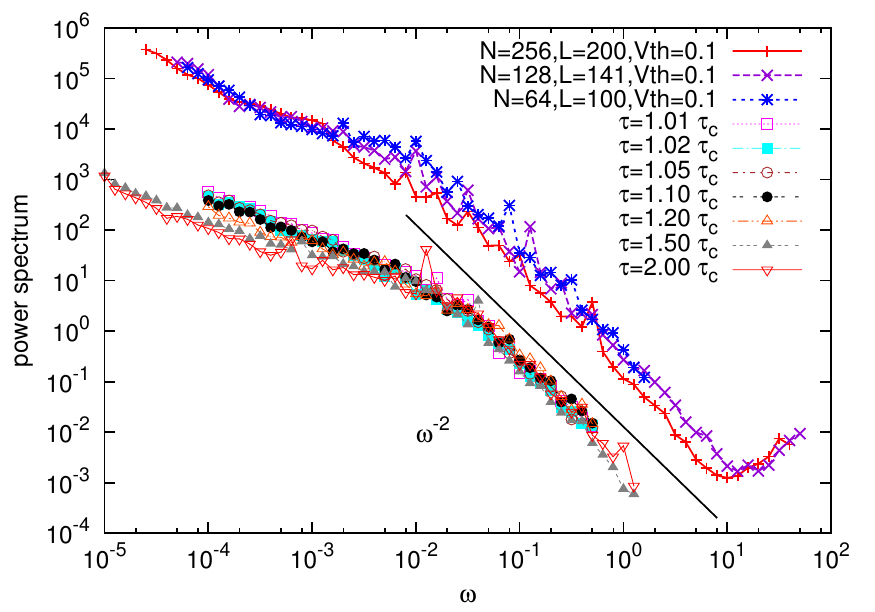}

\caption{(Color online) The power spectrum of the activity due to an
adiabatic increase in the external stress gives a power law of
$\frac{1}{\sigma \nu z} \approx 2$ (top 3 lines). The power-law regime
corresponds approximately to the inverse of the $D_{T}(T)$ power-law
region. At low frequencies (left of shown power-law fit) finite-size
effects truncate the power law. Extracted from 288 runs of the system
with $N=64$ dislocations in a box of $L=100$ and from 96 runs for the
systems with $N=128$ and $L=141$, and $N=256$ and $L=200$. The bottom 7
lines show the power spectra above $\tau_c$ from 96 runs of a system
with $N=64$ and $L=100$. They exhibit the same power law of
$\frac{1}{\sigma \nu z} \approx 2$ (shifted horizontally lower by 100
only to appear separate; all power spectra curves exhibit similar
amount of power).
 %At low frequencies (left of shown power-law fit) finite-size effects come into play
 %(which cannot be cleanly extracted due to the periodic boundary conditions).
 %At low frequencies (left of shown power-law fit) finite-size effects come into play and
%cutoffs appear that should scale as $\omega_{\mm{min}} \sim L^{-\nu z}$. However the periodic boundary
%conditions allow dislocations to travel several times beyond $L$
%and mask that effect.
} \label{PS}
\end{figure}

We calculated the power spectra of the time series of the activity
$V(t)$ for all stresses, i.e., $0<\tau<\tau_{c}$
(integrated-over-stress), using the Lomb periodogram technique
~\cite{NumRecC}. The stress-integrated avalanche size distribution exponent is $\kappa+\sigma=2$. 
As shown by Kuntz and Sethna~\cite{Kuntz00} for a size distribution exponent less than or equal to $2$ the power spectrum scales as
\begin{eqnarray}
PS_{\mm{int}}(\omega) = \bigg | \int{V(t)e^{i \omega t}dt} \bigg |^2 \sim \omega^{-\frac{1}{\sigma \nu z}}.
%sim \omega^{\frac{\kappa+\sigma-3}{\sigma \nu z}} 
\label{PSint}
\end{eqnarray}
Our results are shown in fig.~\ref{PS} where we find $\frac{1}{\sigma \nu z} \approx 2$.

\begin{figure}[!h]
\centering
    \includegraphics[width=0.49\textwidth]{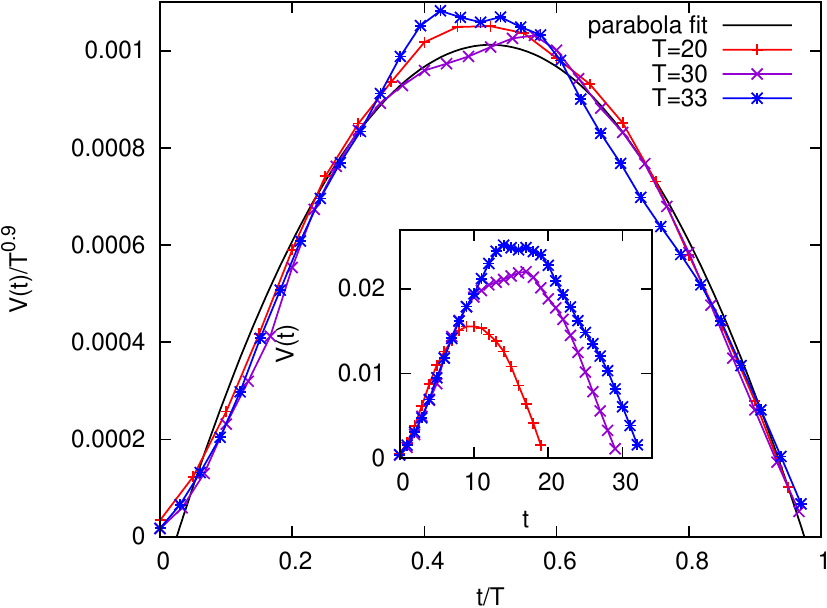}

\caption{(Color online) Scaling collapse of the avalanche shapes
(uncollapsed shapes shown in inset). It gives $\frac{1}{\sigma \nu z}
\approx 1.9$ in agreement with the power spectra. (Inset) Averaged
avalanche profiles (shapes)  for 3 different durations from the
power-law regime of $D_{T}(T)$. Extracted from 96 runs with  $N=64$
dislocations in a box of $L=100$. (Note that $V_{\mm{th}}=0.1$ was
subtracted from the signal $V(t)$).} \label{AvalShapes}
\end{figure}

The duration of an avalanche is $T=t_{\mm{end}}-t_{\mm{start}}$.
From our simulations we extract the avalanche shapes in the pinned
phase. We collect all the avalanches within $\pm5\%$ of a given
duration and average their temporal profiles. For sufficiently small durations
the avalanches are taken from the power law regime of the duration
distribution. We collapse them using~\cite{Kuntz00,Sethna01}
\begin{eqnarray}
V(t)=T^{\frac{1}{\sigma \nu z}-1}f_{\mm{shape}}(t/T).
\label{Vshape}
\end{eqnarray}
We obtain a good collapse, which indicates that the scaling exponent
has the MF value of $\frac{1}{\sigma \nu z} \approx 2$ and the scaling
function $f_{\mm{shape}}$ is a parabola, same as in MFT (fig.
~\ref{AvalShapes}). In addition the power spectra exponent and the
exponent that collapses the avalanche shapes are in excellent agreement
with each other. In~\cite{Laurson06} the avalanche shapes were first
rescaled with an assumed exponent of $1/\sigma \nu z \approx 1.5$ and
then averaged. This is not the same as the Widom  scaling collapse
presented here. We first average shapes of avalanches of the same
duration. Then we tune the exponents until the average shapes of
different durations collapse. In our case the exponent $\frac{1}{\sigma
\nu z} \approx 2$ is a result of the scaling collapse and does not need
to be assumed up front. Also in~\cite{Laurson06} a power spectra
exponent of $1/\sigma \nu z \approx 1.5$ was fitted for the activity
fluctuations above the critical flow stress while the system was in a
steady state. In contrast, our power spectra above the critical flow
stress give the same power-spectra exponent of  $\frac{1}{\sigma \nu z}
\approx 2$ as our power spectra below the critical flow stress
indicating that the critical region extends at least up to $\tau = 2.0
\tau_c$ (see fig.~\ref{PS}).

%In~\cite{Laurson06} an avalanche shape collapse
 %was attempted with a preconceived exponent of $1.5$. In general in the depinned phase
%individual avalanches tend to merge together and their resulting shape
%may be different from the pure single avalanche shape
%and may not be amenable to a collapse. Also in~\cite{Laurson06}
%a power spectra exponent of $1.5$ was
%found for the activity fluctuations above the flow stress while the
%system was in a steady state. In contrast, our power spectra
%above the flow stress give the same power-law exponent of $2$
%as our power spectra below the flow stress indicating that
%the critical region extends at least up to $\tau=2.0\tau_{c}$ (see fig.~\ref{PS}).

We also extract the probability distribution of the avalanche sizes,
durations and energies. We define the size of an avalanche as
$S=\int_{T}V(t)dt$ and the energy as $E=\int_{T}V^{2}(t)dt$.
The distribution of energies at different stresses can be shown to
scale as $D_{E}(E,\Delta) \sim E^{-1-\frac{\kappa-1}{2-\sigma \nu
z}}f_{E}(E\Delta^{\frac{2-\sigma \nu z}{\sigma}})$ and the distribution
of durations as $D_{T}(T,\Delta) \sim T^{-1-\frac{\kappa-1}{\sigma \nu
z}}f_{T}(T\Delta^{\nu z})$ assuming that the distribution of sizes
scales as $D_{S}(S,\Delta) \sim S^{-\kappa}f_{S}(S\Delta^{\frac{1}{\sigma}})$
~\cite{ZaiserYielding}, the correlation length scales as $\xi \sim \Delta^{-\nu}$ and the
dynamic exponent $z$ is defined via $T \sim \xi^{z}$. $\Delta
=1-\tau/\tau_{c}$. All distributions have a power-law region (e.g.
$D_{S}(S,\Delta) \sim S^{-\kappa}$) up to a cut-off or maximum
avalanche (e.g. $S_{\mm{max}} \sim\Delta^{-\frac{1}{\sigma}}$) which
increases as the system approaches the critical flow stress from below. We
calculated all the power-law and cut-off exponents above from our
simulations and they are consistent with MFT. Only
%the power-law exponent, $1+\frac{\kappa-1}{\sigma \nu z}$, of the
$D_{T}(T)$ could not be extracted due to finite-size effects. Also,
much larger system sizes than our maximum of $L=200$ and $N=256$ are
needed to collect enough statistics for the largest avalanches where
the power-law scaling region of the distributions cuts off and measure
the correct stress-integrated ($0<\tau<\tau_{c}$) power-law exponents
(e.g. $D_{S,\mm{int}}(S) \sim S^{-\kappa-\sigma}$).

As the system approaches the critical flow stress from below, $\tau \to
\tau_c$, (i.e., the critical point of the depinning transition) the
correlation length diverges, $\xi \sim (1-\tau/\tau_{c})^{-\nu}$. Up to
$\xi < L$ the maximum avalanche is given by $S_{\mm{max}}
\sim\Delta^{-\frac{1}{\sigma}} \sim \xi^{\frac{1}{\sigma \nu}}$.
However when the correlation length outgrows the system size, $\xi >
L$, the maximum avalanche is dictated by the system size, $S_{\mm{max}}
\sim L^{\frac{1}{\sigma \nu}}$. We can quantify the finite-size effects
through the exponent $\nu$ (fig.~\ref{vtau}). The integrated size
distribution can bevmodified to account for finite-size effects,
$D_{S,\mm{int}}(S,L) \sim S^{-(\kappa+\sigma)}f_{S,\mm{int}}(SL^{-\frac{1}{\sigma
\nu}})$. We were able to qualitatively observe the increase of the
maximum avalanche of $D_{S,\mm{int}}(S,L)$ with $L$.
We quantify that dependence through the moments, $\langle S^{m} \rangle =
\int^{S_{\mm{max}}}_{0}S^{m}D_{S,\mm{int}}(S,L)dS$. For
$m>\kappa+\sigma$ the integral does not diverge at the lower limit and
we get $\langle S^{m} \rangle \sim L^{\frac{1+m-\kappa-\sigma}{\nu
\sigma}}$. By plotting $\mm{log_{10}}(\langle S^{m+1} \rangle/\langle S^{m}
\rangle) \sim \frac{1}{\nu \sigma} \cdot \mm{log_{10}(L)}$ we obtain consistent
values for $\nu$, independent of $m$ (see inset of fig.~\ref{vtau}). Same when using $\langle E^{m} \rangle \sim
L^{(m-\frac{\kappa-2}{2-\sigma \nu z})\frac{2-\sigma \nu  z}{\nu
\sigma}}$. Unfortunately applying
$\langle T^{m} \rangle \sim L^{(m-\frac{\kappa+\sigma-1}{\sigma \nu z})z}$
to the data does not yield reliable results because the durations
are plagued by large finite-size effects and large errorbars.
%We extract the dynamic exponent $z$ from the values of $\nu$ and $\nu z$, and
We present all exponents in Table~\ref{table}.

\section{Above the critical flow stress} 
\begin{figure}[h]
\centering
    \includegraphics[width=0.49\textwidth]{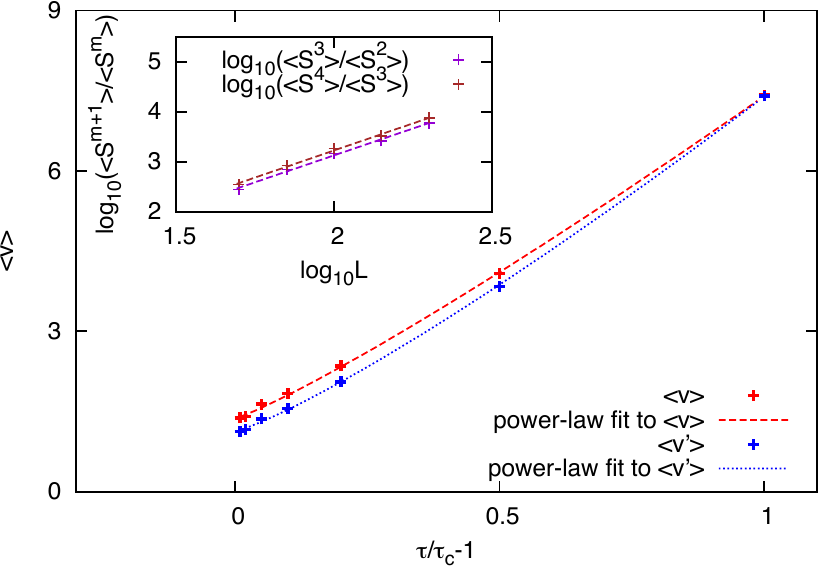}

\caption{(Color online) Mean collective speed $V(t)$ and mean strain
rate $V'(t)$ plotted against the reduced stress above the flow stress.
The power-law fits yield: $\langle V \rangle \sim ( \tau/\tau_{c} - 1 )^{1.14(\pm0.02)}$
and $\langle V' \rangle \sim ( \tau/\tau_{c} - 1)^{1.19(\pm0.02)}$.
%the forms: $\langle V \rangle \approx 1.38(\pm 0.03)+6.03(\pm0.05) ( \tau/\tau_{c} - 1 )^{1.14(\pm0.02)}$
%and $\langle V' \rangle \approx 1.13(\pm 0.03)+6.26(\pm0.04) ( \tau/\tau_{c} - 1)^{1.19(\pm0.02)}$.
%The value of $\beta$ is more sensitive to the number of points used in the fit than the statistical errors indicate.
%The statistical errors quoted are not the dominant errors.
%In the numerical study of a critical phenomenon the systematic errors dominate.
%The estimated systematic errors dominate the statistical errors.
%We obtain $\beta = 1.1 \pm 0.1$, the MF value.
Each of the 7 points ($\tau/\tau_{c}=1.01,1.02,1.05,1.1,1.2,1.5,2.0$)
is extracted from 96 runs with $N=64$ dislocations and $L=100$. The
$\tau/\tau_{c}$ points used exhibit power-law power spectra (fig.
~\ref{PS}) and therefore are part of the critical region. We expect the
systematic error that comes from determining $\tau_c$ to be larger than
the statistical error above. Using exponent relations we get $\beta =
1.0\pm0.2$, consistent with MF.
%the lowest values are plagued by the fact that we make a systematic error in trying to get the exact flow stress right
(Inset) Finite-size scaling analysis for the
 dislocation system at fixed density $N/L^2=16/50^2=32/71^2=64/100^2=128/141^2=256/200^2$. 
 Spanning avalanches were excluded
(a spanning avalanche has at least one dislocation travel by $L$).
The linear fits on the moment ratios (dashed lines) yield $\nu$
using $\mm{log_{10}}(\langle S^{m+1} \rangle/\langle S^{m}
\rangle) \sim \frac{1}{\nu \sigma} \cdot \mm{log_{10}(L)}$.
See Table~\ref{table} for results and text for details.}
\label{vtau}
\end{figure}

The critical flow stress $\tau_c$ for each
system is the stress reached at the end of the adiabatic run. At that
stress we observe the last infinite avalanche with the dislocations
moving out of the basic cell at one side and in at the other for
periodic boundary conditions. The critical flow stress $\tau_{c}$ is not a
universal quantity and every system with the same number of
dislocations and box size has a different $\tau_{c}$. Knowing the
critical stresses from the adiabatic run allows us to simulate at a
fixed fraction above each realization's own critical flow stress. We obtain a
sharp transition from the pinned to the depinned phase and a linear
relationship between mean dislocation activity, $\langle V \rangle$,
and distance from the critical point:
\begin{eqnarray}
\langle V \rangle \sim (\tau/\tau_c-1)^{\beta} \hspace{2mm} \mm{with} \hspace{2mm} \beta \approx 1
\label{beta}
\end{eqnarray}
(fig.~\ref{vtau}). This result agrees with MFT predictions
~\cite{Zapperi98,Fisher97, UhlPRL09}, but differs from Ref.
~\cite{MiguelPRL02} where the critical flow stress was determined in a
collective manner for the entire ensemble. Our treatment properly
accounts for the ensemble stress fluctuations~\cite{TsekenisJam11} on
$\beta$, but we expect that the two approaches should yield the same
exponent in the thermodynamic limit.

We also calculated the power spectra at fixed stress above the critical flow stress.
The stress-binned avalanche size distribution exponent is $\kappa<2$.
This means that the power spectra at fixed stress scale the same 
as $PS_{\mm{int}}(\omega)$~\cite{Kuntz00}
\begin{eqnarray}
PS(\omega) = \bigg | \int{V(t)e^{i \omega t}dt} \bigg |^2 \sim \omega^{-\frac{1}{\sigma \nu z}}.
\label{PSint}
\end{eqnarray}
In fig.~\ref{PS} we show that the power spectra at fixed stress in the depinned phase ($\tau>\tau_{c}$) 
exhibit identical scaling as the power spectra integrated over all stresses in the pinned phase ($0<\tau<\tau_{c}$).

\section{Discussion} We demonstrated that not just the static but also
the {\it dynamic} characteristics of crystalline deformation, (i.e.,
critical exponents and scaling functions), in the absence of hardening,
belong to the universality class of the mean field (MF) interface
depinning transition. (\lq\lq Absence of hardening"  refers to systems
without dislocation creation and annihilation, and with a zero {\em
slope} of the stress-strain curve in the vicinity of the critical flow
stress.) Specifically, we showed that the temporal profiles of the
avalanche shapes collapse on to a parabolic MF scaling function with a
MF scaling exponent, $\frac{1}{\sigma \nu z}=2$. This value agrees, as
predicted, with the scaling exponent of the power spectrum of the
acoustic emission signal during plastic deformation. We provided a
finite-size scaling analysis of dislocation systems that shows the
value of the finite-size scaling exponent $\nu$ is also consistent with
MFT predictions. We extracted the depinning exponent $\beta$ which
characterizes the dynamic interface depinning phase transition by
taking proper care of the ensemble fluctuations and found it in
accordance with MFT . Our work thus resolves the differences between
prior results on static and dynamic plasticity exponents, and shows
that both static and dynamic exponents and scaling functions belong to
the MF interface depinning universality class~\cite{UhlPRL09}.

\acknowledgments
We thank M.-C. Miguel, M. Zaiser, J. Weiss, S.
Zapperi, L. Laurson, M. Alava, D. Ceperley, V. Paschalidis, Y.
Ben-Zion, K. Schulten, T. M. Earnest, N. Friedman, A. Jennings, J. Greer and J. Sethna for helpful
conversations. We acknowledge NSF grant DMR 03-25939 ITR (MCC) and DMR
10-05209, SCEC, the University of Illinois Taub cluster and NSF grant
TG-DMR090061 for TeraGrid TACC and NCSA resources.

\bibliographystyle{eplbib}
\bibliography{one}

\end{document}